# Detecting "protein words" through unsupervised word segmentation

Wang Liang[1], Zhao KaiYong[2]

[1]Sogou Tech, Beijing, 100080, P.R. China.

[2]Department of Computer Science, Hong Kong Baptist University, Hong Kong, 999077, P.R. China.

Correspondence to Wang Liang: wangliang.f@gmail.com

## ABSTRACT

Unsupervised word segmentation methods were applied to analyze protein sequences. Protein sequences, such as "MTMDKSELVQKA…," were used as input to these methods. Segmented "protein word" sequences, such as "MTM DKSE LVQKA," were then obtained. We compared the "protein words" derived via unsupervised segmentation and protein secondary structure segmentation. An interesting finding is that unsupervised word segmentation is more efficient than secondary structure segmentation in expressing information. Our experiment also suggests the presence of several "protein ruins" in current non-coding regions.

## Introduction

Word segmentation mainly refers to the process of dividing a string of written language into its component words. For some East Asian languages such as Chinese, no space or punctuation is placed between letters. The "letter" sequences must be segmented into word sequences to process the text of these languages. For example, "iloveapple" is segmented into "I love apple."

The main concept of the segmentation method is simple. For example, the letter sequence "iloveapple" can be segmented into many forms of word sequences, including "I love apple," "ilove apple," "il ove apple," "iloveapple," "ilo vea pple," etc. We select the "word sequences" with the highest probability as the segmentation of this letter sequence. For example,

P("I love apple") = 0.8,

P("ilove apple") = 0.3,

P("iloveapple") = 0.03.

Hence, we will select "I love apple" as the segmentation of letter sequence "iloveapple."

The probability of a word sequence is calculated by the probability of the multiplication of each word.

For example,

P("I love apple") = P(i) ×P(love) ×P(apple),

P("ilove apple") = P(ilove) ×P(apple).

Therefore, the key to segmentation is determining the probability of each word.

If we have many possible segmented sequences such as that for "i love apple," then we can simply calculate the probability of each word by dividing the number of its occurrences by the number of occurrences of all the words. For example, we have two sequences, namely, "i love apple" and "i love mac;" that is, we have six word occurrences. "i" appears two times; thus, its probability P("i") = 2/6. "apple" appears one time; thus, its probability P("apple") = 1/6. This method is called supervised segmentation. For words such as "ilove" that do not appear in corpus, an extremely small probability is provided.

**If we have no segmented sequence but many letter sequences such as "iloveapple," then we can still calculate the probability of all the possible words and develop the word segmentation method.** The main concept of this method is described as follows.

First, the random or equal probability of each possible word is provided. In general, we need to set a maximal word length. For example, for "iloveapple," if we set four as the maximal word length, then possible words include "ilov," "love," "ovea," etc.

Second, given that we have the word list with probability, we can segment all the letter sequences into "word sequences."

Third, we calculate the new word probability according to the "word sequences" in the second step.

We repeat the second and third steps until the probabilities of all the words reach a stable value. Then, we can use this word list with probability to segment letter sequences. This method is called unsupervised segmentation [1].

The basic theory of word segmentation is depicted in Fig.1.

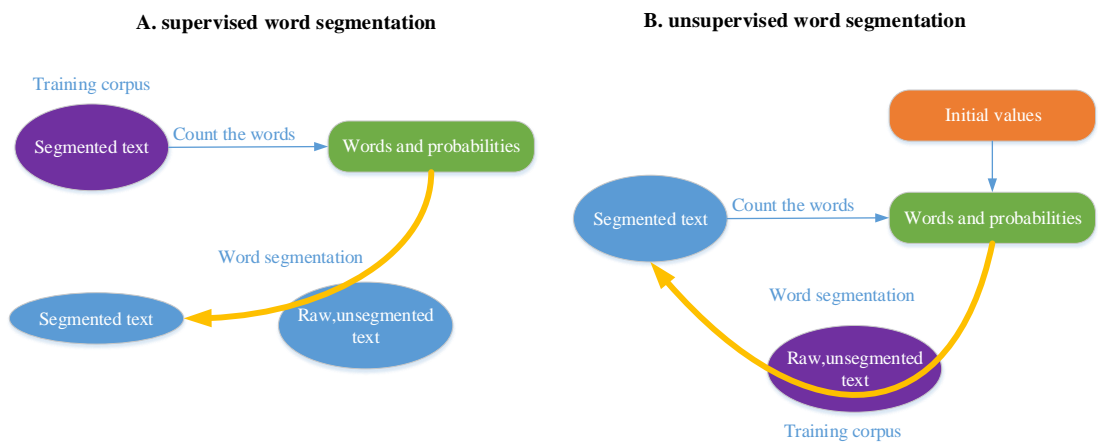

Fig.1. A. Supervised word segmentation. B. Unsupervised word segmentation.

For English text, the accuracy of supervised segmentation is over 95%. Meanwhile, the unsupervised segmentation method can reach approximately 75%. Supervised methods are better than unsupervised methods; however, unsupervised approaches are highly adaptive to relatively unfamiliar "languages" for which we do not have sufficient linguistic knowledge [2].

Unsupervised segmentation can be described as "inputting letter sequences and outputting meaningful word sequences." If the input is a protein sequence such as "MTMDKSELVQKA…," then what is the output? The answer to this question is the topic of this study.

In the following sections, we will perform an unsupervised segmentation of protein sequences. Identifying the functional equivalents of "words" in protein sequences is a fundamental problem in bioinformatics. The elements of protein secondary structure are generally regarded as the functional building blocks of protein sequences. Therefore, we compare our "protein word" sequences and related secondary structure segmentation in Section 3. We find our "protein word segmentation" to be a better encoding method than secondary structure segmentation in terms of expressing information. We discuss this confusing result in Section 4. The last section provides a short summary of the study.

## Word segmentation experiment

The soft-counting method is a representative unsupervised segmentation method; it uses the standard expectation–maximization algorithm to estimate the probability of any possible substring [1]. This operation produces a vocabulary with probability. To segment a sequence, a Viterbi-style algorithm is then applied to search for the segmentations with the highest probabilities.

Unsupervised segmentation methods require only raw letter sequences. In this study, we use amino acid letter sequences. Data are selected from the Protein Data Bank (PDB) structure database. The PDB database contains approximately 100,000 protein sequences. These sequences are used as input in the unsupervised segmentation method. Thus, we obtain

**Definition 1**: **"protein word,"** the segments produced by a word segmentation method.

Unsupervised segmentation only requires selecting a maximal word length. In this study, we set nine as the maximal protein word length. This value can be adjusted according to data size.

Approximately 26 million possible protein words are included in our data. Frequency and border information features are used to filter these words [3–6]. After performing the soft-counting method for these protein sequences, we obtain a final protein vocabulary with approximately 630,598 words. We can then use this vocabulary with word probability to segment protein sequences. An example of protein word segmentation is shown as follows.

Original protein sequence:

**DADLAKKNNCIACHQVETKVVGPALKDIAAKYADKDDAATYLAGKIKGGSSGVWG QI PMPPNVNVSDADAK LADWILTLK**

Corresponding segmented protein word sequence:

**D ADLAKK NNCI ACH QVET KV VGPAL KDIAAK YADK D DAATYL AGKIK GGSSGV WGQI PMPPN VNVSD ADAKA LADWI LTLK**

## Comparison with secondary structure segmentation

In most human language texts, the output of segmentation is a meaningful word sequence. As expected, the same is true for a protein sequence. We compare a protein word sequence with the elements of protein secondary structure, which act as the functional building blocks of a protein.

Protein secondary structure mainly refers to the local conformation of the polypeptide backbone of proteins that is frequently and discretely classified into a few states. Word segmentation is highly similar to the protein secondary structure assignment process. For protein sequences, such as "MTMDKSELVQKA," the corresponding consecutive amino acids of the same secondary structure can also be regarded as a "structure word." This process can be called secondary structure segmentation (Fig.2). A total of 437,537 distinct "structure words" are included in our data set.

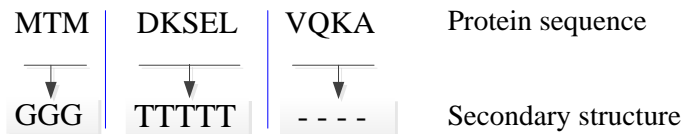

Fig.2. Secondary structure segmentation. A protein sequence is segmented by its secondary structure. The sequence in this figure has three secondary structure words: "MTM," "DKSEL," and "VQKA."

**Definition 2**: **"structure word,"** the protein segments produced by secondary structure segmentation.

Segmentation performance is generally evaluated using the boundary F-score measure, F = 2RP/(R + P). Recall R is the proportion of the correctly recognized word boundaries to all boundaries in the gold standard; precision P is an output of the word segmentation of a segmenter. In this study, we use structure segmentation as the gold standard (Fig.3).

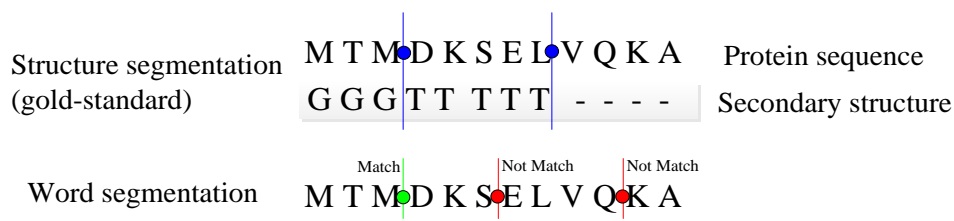

Fig.3. Evaluation of word segmentation. For this example, the precision is 1/33 ≈ 3% and recall is 1/2 = 50%. The F-score is 2 ×0.33 ×0.5/(0.33 + 0.5) ≈ 39.8%.

For the soft-counting method, boundary precision is 39.9%, boundary recall is 28.0%, and the boundary F-score is 32.9%. A segmentation example is provided in Fig.4.

```
MVLS EGEWQLVLHVWAKV EAD VAGHGQDILIRLFKS H PETLEK  F DRVKH L   Structure segmentation
- - - - HHHHHHHHHHHHHH GGG HHHHHHHHHHHHHH - GGGGGG -  TTTTT  -  Secondary structure
MV LSEGE WQLV LHVW A KVEAD VAGH GQDIL IRLF K SHPET LEKFD R VKHL  Soft-counting word segmentation
```

Fig.4. Structure segmentation and word segmentation of a protein sequence.

The boundary F-score is approximately 85% for unsegmented English text using the soft-counting method. For Chinese text, the boundary F-score is approximately 80%. For protein word segmentation, however, such value is only 32.9%. Therefore, protein word segmentation is different from secondary structure segmentation. This result does not satisfy our expectation. We will analyze the differences in the next section.

## Encoding efficiency of segmentation

As a basic statistical feature, word occurrence distribution may explain the differences between the two kinds of segmentation.

In a typical English data set, if we regard the top 10% frequency words in vocabulary as high-frequency words, then the occurrences of these words generally account for approximately 90% of all the letters in the data set. For low-frequency words, for example, 1-frequency (only appear once) words account for approximately 50% of all the words in vocabulary; however, their occurrences only account for approximately 1% of the letters of the entire corpus. This law is "efficient" for most human languages. It is also a fundamental assumption in most unsupervised segmentation methods. The related values of soft-counting segmentation and secondary structure segmentation are provided in Table 1.

Table 1: Word occurrence distribution

|  | Letter percentage of high-frequency words in the data set | Letter percentage of low-frequency words in the data set |
|---|---|---|
| English text | 80% | 3% |
| Soft-counting segmentation | 34% | 4% |
| Structure segmentation | 40% | 58% |

Based on Table 1, secondary structure segmentation appears to be "inefficient" primarily because its vocabulary contains too many low-frequency words. We can use "description length (DL)" to describe the "efficiency" of segmentation [7]. The segmentation process can be regarded as an encoding process that replaces the letters of a word with related word symbols. A codebook, in which each word is represented by a unique string, can be used to encode a corpus of words as efficiently as possible. The total number of letters required to encode the corpus (i.e., the sum of the lengths of the codebook and the encoded corpus) that uses a well-designed codebook/vocabulary will be less than that for the original corpus. Smaller units, such as morphemes or phonemes, which require fewer code words, and thus, a shorter codebook, can be encoded further. However, efficiently encoding the corpus becomes increasingly difficult as the number of code words used decreases. Meanwhile, some words may never be used when the codebook has too many words. Thus, the length of the codebook and the length of the encoded corpus must be balanced. The DL principle states that a codebook that results in the shortest total combined length must be selected. Therefore "DL" can be used to describe segmentation efficiency.

The DL values of structure segmentation and soft-counting segmentation are provided in Fig.5.

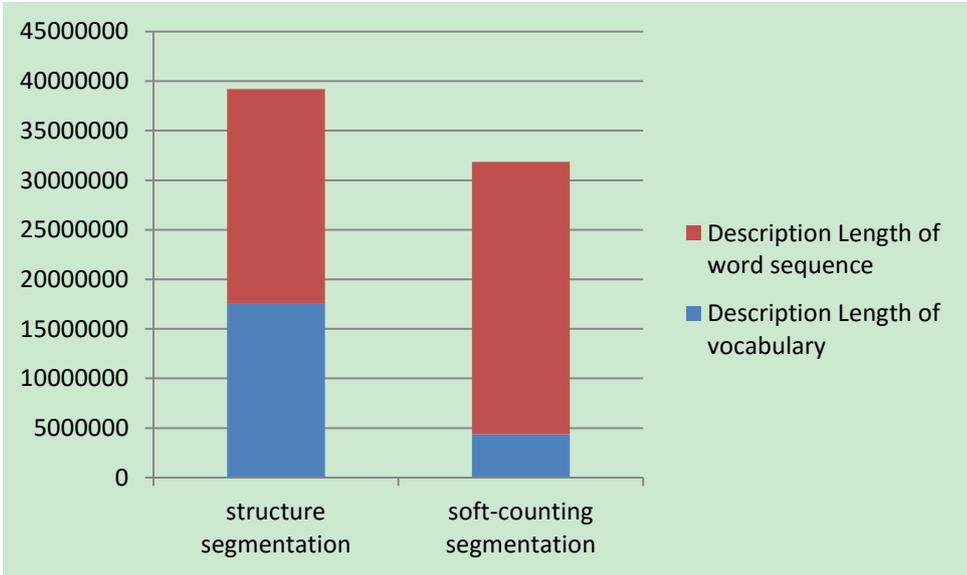

Fig.5. DL values of structure segmentation and soft-counting segmentation.

From Fig.5, we can determine that the entire DL of structure segmentation is more than that of soft-counting segmentation. In structure segmentation, approximately 45% of the DL value is used for vocabulary. In soft-counting segmentation, such value is only 16%. Structure segmentation designs a large "codebook," but only a few of its capabilities are used. Thus, secondary structure segmentation may not be a good encoding method. Two opposing explanations may explain this result.

First, current structure segmentation is "really" inefficient. Therefore, we can find a different structure assignment method whose structure segments can express the functions of genes better. Our data set uses the DSSP secondary structure definition. We also try other kinds of secondary structure definitions such as STRIDE [8], but we obtain similar low "efficiency" segmentation.

Second, the structure segmentation scheme is "truly" efficient. Given that the segmentation method is appropriate, some problems may exist in the data set. A highly common problem in the data set construction process is the unbalanced data problem. For example, if we only select several paragraphs of a long book to build a data set, then this problem will occur.

We develop the experimental data by filtering similar protein sequences in PDB. The PDB protein data can also be regarded as randomly selected from all available protein data. No obvious problem is observed in our data construction process. Hence, if our data set is unbalanced, then all current protein data are unbalanced. If yes, then more "protein words" are present in current non-coding regions. For example, several short DNA regions can be translated into "protein words;" however, such "protein words" that cover a region may not comply with the strict definition of a "protein" sequence. Therefore, we have

**Definition 3**: **"protein word coverage region,"** the DNA region that can be translated into "protein words."

We can also detect a "protein word coverage region" using segmentation methods. We back-translate the protein words into DNA forms and obtain a DNA vocabulary. Then, we can use this vocabulary to segment DNA sequences using supervised segmentation methods. For example, a DNA sequence "ATGGTTTTGTGTCTTGAGGGT" is segmented into

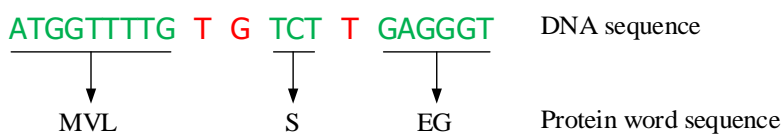

Fig.6. Segment DNA sequence based on a protein word vocabulary. The green parts can be translated into protein words. The red parts cannot be translated. (Note: This sequence is not a protein sequence. Only some regions can be covered by "protein words").

The following is a typical example of DNA sequence segmentation:

TTGTGG G TTGTCTATTGATGTT TTTGGT C TATTCTAAG A A TTGGAG A GAG A GAGGTT A A AAT C TCT G ACT ATG A TTGTGG A TTGTCT GCTGAT GTT TTTGGT C TATTGTTCT AAGAAT TGG A GAG A GAG A GAGGTT A A A A TCT C C G ACT ATG A TTGTGG A TTGTCT ATTGCTGTTTTT GGT C TATTGTTCT AAGAAT.

This sequence is part of "gi|157704448|ref|AC_000133.1| *Homo sapiens* chromosome 1, alternate

assembly HuRef, whole genome shotgun sequence:307501-308000." The green segments can be translated into "protein words." Although 87% of the region of this sequence is a protein word coverage region, no distinct intron, exon, and stop codon is found. Hence, such sequence cannot be regarded as a coding sequence.

We use the human genome to judge whether the data imbalance can be improved by adding all the "protein word coverage regions." The genome is divided into 500 bps equal-length sequences. We use two vocabularies, namely, soft-counting word vocabulary and secondary structure word vocabulary, to segment the aforementioned DNA sequences. Then, we calculate their DL individually. The results are presented in Fig.7.

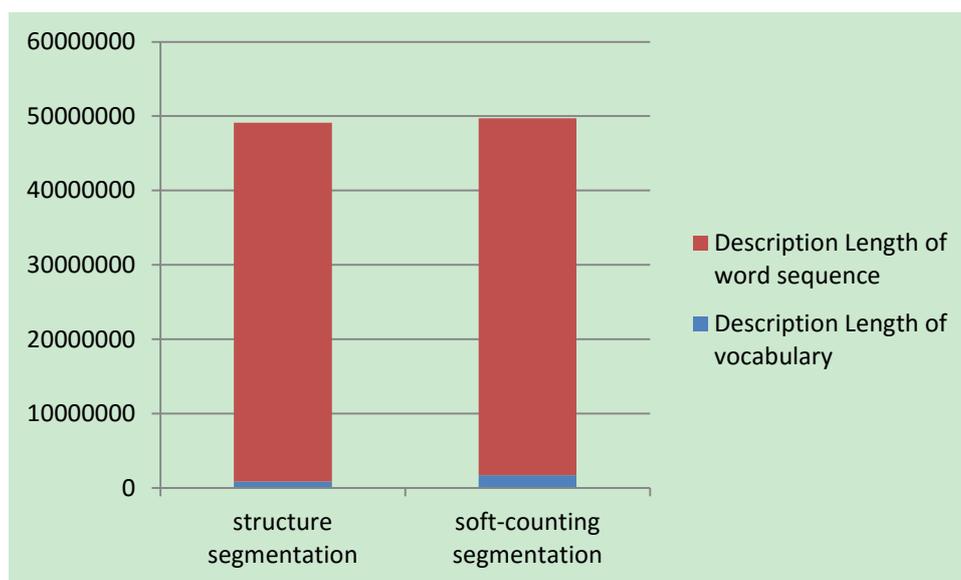

Fig.7. DL values of structure segmentation and soft-counting segmentation.

We determine that the two kinds of segmentation methods have similar DL values. Then, we convert all the DNA "protein word coverage regions" into protein word sequences. These sequences are regarded as gold standard segmentation. Then, we delete the space between words and perform unsupervised segmentation test again. Boundary precision reaches 67%, boundary recall reaches 60%, and the boundary F-score reaches 63%. These results concur with our conjecture. Adding "protein word coverage regions" can address the data imbalance and improve structure segmentation efficiency.

The distribution of "protein word coverage regions" is shown in Fig.8.

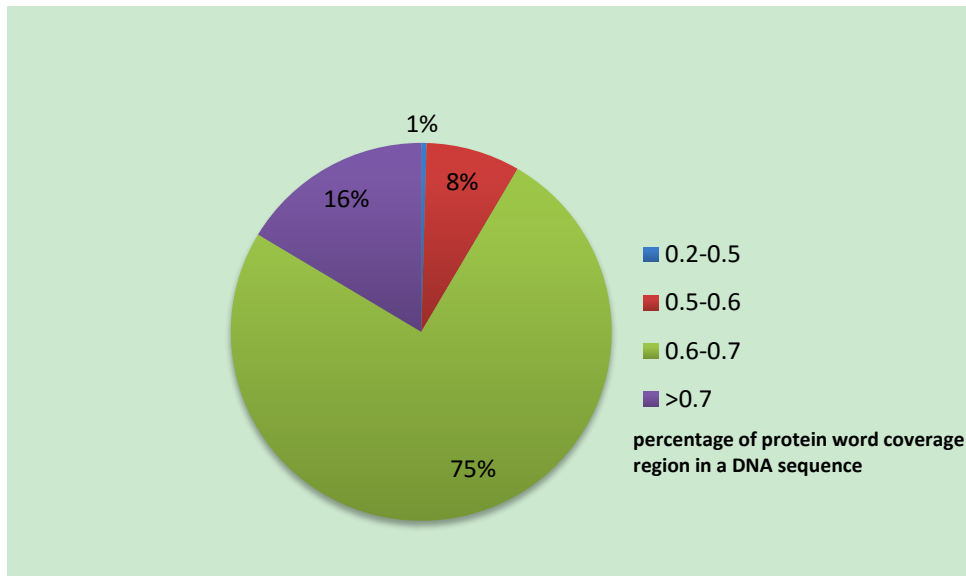

Fig.8. Distribution of the protein word coverage sequence in *Homo sapiens* chromosome 1. For example, DNA sequences whose percentage of protein word encoding region is within [0.6, 0.7] are approximately 75%.

We determine that approximately 16% of the DNA sequences have a "protein word coverage region" percentage of over 70%. Among these sequences, only approximately 10% are actual gene areas, the remaining sequences may be regarded as "protein ruins," which may be abandoned in evolution.

## Conclusion

We use protein sequences as input to perform a word segmentation test. As expected, the output is several meaningful "protein word" sequences. Thus, we compare them with secondary structure segments, which are generally regarded as the basic functional blocks of a protein. However, we determine that the two kinds of segmentation methods are different. Subsequently, we use "DL" to analyze their difference and determine that word segmentation is more efficient than structure segmentation. This result is irrational. We suppose that this problem is mainly caused by the data imbalance. That is, the gene is similar to a large book and current "protein" sequences are only several paragraphs in this book. Our DNA segmentation experiments also support this conjecture.

Overall, "protein word" and "protein word coverage region" may have no biological meaning. Word segmentation is a fundamental technology in processing Chinese and other languages without space or punctuation between words, similar to that of protein sequences. Word segmentation is the preliminary step in search engine systems, translation systems, and proofreading systems, etc. The main objective of this study is to inspire new ideas that apply these technologies in analyzing protein sequences. Word segmentation simply acts as a bridge.

# Acknowledgments

# MATERIALS AND METHODS

## Protein Data

Unsupervised segmenting methods only need raw letter sequence. We mainly use the data of PDB (http://www.rcsb.org/pdb/ ) as our experiment data. This dataset contains about 100,000 pieces of protein sequences. We use CD-HIT algorithms to delete the similar protein sequence. Its codes could be found in : http://weizhong-lab.ucsd.edu/cd-hit/download.php .

## Method and source code

All source codes and experiment instructions of this paper could be found in:
https://github.com/maris205/secondary_structure_detection